\documentclass[12pt]{article}
\usepackage{amsmath, amssymb, mathrsfs, mathtools, tensor, braket}
\usepackage{xcolor, graphicx, subfigure}

\usepackage{cite, varioref}

\usepackage{enumitem, float}

\usepackage{caption}
\DeclareCaptionFormat{myformat}{#3}
\evensidemargin \oddsidemargin

\begin{document}

\begin{titlepage}

\begin{center}
\vspace{5mm}
    
{\Large \bfseries
Field Theory of Superconductor and Charged Vortex 
}\\[17mm]
Yoonbai Kim,
\quad SeungJun Jeon,
\quad Hanwool Song
\\[3mm]  
{\itshape
Department of Physics,
Sungkyunkwan University,
Suwon 16419,
Korea
\\[-1mm]
yoonbai@skku.edu,~
sjjeon@skku.edu,~
hanwoolsong0@gmail.com
}
\end{center}
\vspace{15mm}

\begin{abstract}
A Lagrangian of a Schr\"{o}dinger type complex scalar field of Cooper pair, a U(1) gauge field of electromagnetism, and a neutral scalar field of acoustic phonon with constant background charge density is proposed for an effective field theory of conventional superconductivity. We find static charged vortex solutions of finite energy and, for the critical couplings of the quartic self-interaction coupling of complex scalar field and the cubic Yukawa type coupling between neutral and complex scalar field, these charged vortices saturate the BPS (Bogomolny-Prasad-Sommerfield) bound, that guarantees the nonperturbative classification of type I and I$\!$I superconductors.
\end{abstract}

\end{titlepage}

It has been 70 years since the Ginzburg-Landau theory \cite{Ginzburg:1950sr} explained type I and I$\!$I superconductors and vortices\cite{Abrikosov:1956sx}. However,
it seems unlikely to identify the Lagrangian of corresponding effective field theory in consensus \cite{Polchinski:1992ed,Shankar:1993pf,Nagaosa:1999ud,tinkham2004introduction, Wen:2004ym, Bennemann:2008, Coleman:2015,Burgess:2020tbq, Arovas:2019}.
In this work we propose an effective field theory of a complex scalar field of Cooper pair, a U(1) gauge field of electromagnetism, and a neutral scalar field of acoustic phonon with constant background charge density and test some properties  of conventional superconductivity.

If we begin with the Abelian Higgs model, the Ginzburg-Landau free energy is identified as the energy of static configurations. Then the superconducting vacuum and the neutral Abrikosov-Nielsen-Olesen vortices with classification of type I and I$\!$I superconductors are satisfactorily derived
\cite{Abrikosov:1956sx,Nielsen:1973cs,Bogomolny:1975de}. Despite of this success, the model is unsatisfactory because the Cooper pair below critical temperature of the order of Kelvin is depicted by a relativistic complex scalar field. If $\phi$ becomes nonrelativistic just by replacing the light speed $c$ to a slow propagation speed $v_{{\rm p}}$ $(v_{{\rm p}}\ll c)$, all the aforementioned static properties are reproduced automatically but  we are left with a task to verify such characteristic speed in experiments
\cite{Kim:2024gfn,Jeon:2025snd}. The systematic way is to take nonrelativistic limit of the Abelian Higgs model by $\phi = e^{-\frac{i}{\hbar} m c^{2}} \Psi$. The obtained Lagrangian density involves a relativistic U(1) gauge field $A^{\mu}=(\Phi/c,A^{i})$ and a nonrelativistic Schr\"{o}dinger type complex scalar field $\Psi$ of Cooper pair,
\begingroup
\allowdisplaybreaks
\begin{align}
\mathscr{L}_{0}
=
- \frac{ \epsilon_{0} c^{2} }{4}
F_{\mu\nu} F^{\mu\nu}
+ \frac{i \hbar }{2} (\bar{\Psi} \mathcal{D}_{t} \Psi 
- \overline{\mathcal{D}_{t} \Psi} \Psi ) 
- \frac{\hbar^{2}}{2m} \overline{ \mathcal{D}_{i}\Psi}\mathcal{D}_{i}\Psi
-\lambda (|\Psi|^{2} - v^{2})^{2}
,
\label{001}
\end{align}
\endgroup 
where $
\mathcal{D}_{t} \Psi
=
(
\partial_{t}
+
i \frac{q}{ \hbar} \Phi
) \Psi
$ and $
\mathcal{D}_{i} \Psi
=
(
\partial_{i}
-
i \frac{q}{\hbar} A^{i}
) \Psi$ denote gauge-covariant derivative. Finite energy configuration requires the scalar amplitude $|\Psi|$ to obey  
$\displaystyle{\lim_{|\boldsymbol{x}|\rightarrow\infty}|\Psi|=v}$, but it leads to infinite charge due to spatial volume $V$, $Q=q\int dV\,|\Psi|^{2}\sim qv^{2}V$. In order to tame the infinite charge in superconducting sample, it is natural to introduce a constant background charge density $qn_{{\rm s}}$ which couples to the scalar potential as 
$qn_{{\rm s}}\Phi$. Then the charge density includes background charge as $\rho=q(|\Psi|^{2} - n_{\rm s})$.  The superconducting vacuum also called the Higgs vacuum, $\Psi=v$, is given by the configuration of minimum zero energy with the help of vanishing electric and magnetic field. The relation between vacuum expectation value $v$ and background density $n_{{\rm s}}$, $v^{2}=n_{{\rm s}}$, is derived from electrical neutrality of the vacuum $\boldsymbol{E}={\bf 0}$ that explains perfect conductivity. Even with constant external magnetic field, the vacuum of minimum zero energy is achieved by perfect cancellation of the external magnetic field called the Meissner effect, which leads to perfect diamagnetism. Once vortex solutions of vorticity $n$ are investigated with cylindrical symmetry for simplicity, 
$\Psi=|\Psi|(r)e^{-in\theta}$, the nontrivial profiles of scalar amplitude satisfying the boundary conditions, $|\Psi|(0)=0$ and 
$\displaystyle{\lim_{r\rightarrow\infty}|\Psi|=v}$, let the charge nonzero. The Gauss' law with this nonzero charge develops a nontrivial scalar potential of logarithmic divergence at large $r$, $\Phi\sim\ln r$. Substitution of this scalar potential into the equation of scalar amplitude does not allow a regular vortex solution because of the last coupling term between scalar potential and scalar amplitude,
\begin{align}
\frac{d^{2} |\Psi|}{dr^{2}} + \frac{1}{r} \frac{d|\Psi|}{dr} 
& = \frac{(A+n)^{2}}{r^{2}} |\Psi| 
+ \frac{1}{2\xi^{2}} \Big(\frac{|\Psi|^{2}}{n_{\rm s}} - 1\Big) |\Psi| + \frac{q}{4\lambda n_{\rm s}\xi^{2} }  
\Phi |\Psi|
,
\label{463}
\end{align} 
where $A(r)$ is the angular component of U(1) gauge field, 
$A^{i}=-(\hbar/q)\epsilon^{ij}x_{j}(A/r^{2})$, $(i,j=1,2)$, and $\xi$ is the correlation length for the Higgs field, 
$\xi =\hbar/2\sqrt{2 m\lambda n_{\rm s}}$. Since the field theory whose dynamics is governed by $\mathscr{L}_{0}$ \eqref{001} or $\mathscr{L}_{0}+qn_{{\rm s}}\Phi$ cannot support the vortices as static regular solutions of finite energy, both candidate theories seem to be inappropriate as a field theory version of the Ginzburg-Landau theory including the Abrikosov-Nielsen-Olesen vortices. Though there is no legitimate effective field theory for conventional superconductivity until now, there are at least a few guidelines to be passed by any candidate field theory such as the 
superconducting vacuum of zero energy at classical level, existence of 
topological vortices as static regular solutions, and the classification of type I and I$\!$I superconductors.

Based on the Lagrangian density $\mathscr{L}_{0}+qn_{{\rm s}}\Phi$, we take into account an acoustic phonon described by a gapless neutral scalar field of propagation speed $v_{N}$, which is interacting with the complex scalar field $\Psi$ of Cooper pair in terms of a cubic Yukawa type interaction of coupling constant $g$,
\begingroup
\allowdisplaybreaks
\begin{align}
\mathscr{L}
=
& 
- \frac{ \epsilon_{0} c^{2} }{4}
F_{\mu\nu} F^{\mu\nu}
+ \frac{i \hbar }{2} (\bar{\Psi} \mathcal{D}_{t} \Psi 
- \overline{\mathcal{D}_{t} \Psi} \Psi ) 
- \frac{\hbar^{2}}{2m} \overline{ \mathcal{D}_{i}\Psi}\mathcal{D}_{i}\Psi
+ q n_{\rm s} \Phi
\nonumber\\
&
+ \frac{1}{2v_{N}^{2}}(\partial_{t}N)^{2}
-\frac{1}{2}(\partial_{i}N)^{2}
-\lambda (|\Psi|^{2} - v^{2})^{2}-g  N(|\Psi|^{2} - v^{2})
.
\label{201}
\end{align}
\endgroup
The gapless neutral scalar field of acoustic phonon has the classical vacuum at equilibrium without displacement $N=0$ as usual for sound waves. Thus the effective field theory of $\mathscr{L}$ \eqref{201} shares the same superconducting vacuum of $|\Psi|=\sqrt{n_{\rm s}}$ as that of 
$\mathscr{L}_{0}+qn_{{\rm s}}\Phi$ with various properties, e.g. zero energy and charge, except for the flat direction along the neutral scalar field.

About topological excitations, we focus on the vortices in the BPS (Bogomolny-Prasad-Sommerfield) limit \cite{Bogomolny:1975de}. Superconducting samples are assumed to be symmetric along the $z$ axis and so are the vortex strings. Hence $z$ dependence of the fields disappears, electric field has two nonzero planar components $\boldsymbol{E} = (E^{1} , E^{2})$, and magnetic field has single component perpendicular to $xy$ plane, $B=B^{3}$, in order  to lower the energy. When the two couplings $\lambda$ and $g$ have critical values
\begingroup
\allowdisplaybreaks
\begin{align}
\lambda=\lambda_{\rm c} &= \frac{\hbar^{2} q^{2}}{8\epsilon_{0} m^{2} c^{2}} 
,\label{312}
\\
g=g_{\rm c} &= \frac{q}{\sqrt{\epsilon_{0}}}
,\label{327}
\end{align}
\endgroup
the energy per unit length along $z$ axis is reorganized by use of the Bogomolny trick \cite{Bogomolny:1975de}, 
\begingroup
\allowdisplaybreaks
\begin{align}
\bar{E} =&\, \int d^{2} \boldsymbol{x}\, \bigg\{  
- \sqrt{\epsilon_{0}} (N + \sqrt{\epsilon_{0}} \Phi) 
\Big[ \boldsymbol{\nabla}\cdot\boldsymbol{E} 
- \frac{q}{\epsilon_{0}} (|\Psi|^{2} - n_{\rm s}) \Big]  
+ \nabla\cdot \big[ \sqrt{\epsilon_{0}} \boldsymbol{E} (N + \sqrt{\epsilon_{0}} \Phi) \big] 
\nonumber\\
&\,~~~~~~~~~~ + \frac{1}{2} (\nabla N - \sqrt{\epsilon_{0}}\boldsymbol{E})^{2} + \frac{\hbar^{2}}{2m} |(\mathcal{D}_{1} \mp i \mathcal{D}_{2}) \Psi|^{2} + \frac{\epsilon_{0} c^{2}}{2} \bigg[ B \mp \frac{\hbar}{2q\lambda_{\rm L}^{2}} \bigg( \frac{|\Psi|^{2}}{n_{\rm s}} - 1 \bigg) \bigg]^{2}  \nonumber\\
&\,~~~~~~~~~~
\pm \partial_{i} (\epsilon^{ij} j_{j})\mp \frac{\hbar q}{2m} n_{\rm s} B \bigg\}
\label{424}
\\
&\,\ge  \frac{\hbar^{2} n_{\rm s}}{2m}
\bigg| \frac{\Phi_{B}}{\Phi_{\rm L}} \bigg|
= \frac{\pi \hbar^{2} n_{\rm s} }{m} |n|
= \frac{\hbar^{2} n_{\rm s}}{4 \epsilon_{0} m^{2} c^{2}}
| q \bar{Q}_{\mathrm{U}(1)} |
,\label{421}
\end{align}
\endgroup
where $\Phi_{B}=\oint_{\partial \mathbb{R}^{2}} d \boldsymbol{l}\cdot\boldsymbol{A}
$ is the magnetic flux, $\Phi_{{\rm L}}=\hbar/q$ is the London flux quantum, and $\bar{Q}_{{\rm U(1)}}= q \int d^{2} \boldsymbol{x} \,\rho
= q \int d^{2} \boldsymbol{x} \,
(|\Psi|^{2} - n_{\rm s})$ is the charge per unit length along $z$ axis. Note that the total divergence term involving current density $j^{i}
=- i \frac{q \hbar}{2m} \big(
\bar{\Psi} \mathcal{D}_{i} \Psi
- \overline{\mathcal{D}_{i} \Psi} \Psi
\big)
$ does not contribute to the energy per unit length along $z$ axis. If the following Bogomolny equations hold, 
\begingroup
\allowdisplaybreaks
\begin{align}
\boldsymbol{\nabla} N - \sqrt{\epsilon_{0}} \boldsymbol{E}
&=0
,\label{422}
\\
(\mathcal{D}_{1} \mp i \mathcal{D}_{2}) \Psi
&=0
,\label{425}
\\
B \mp \frac{\hbar}{2q\lambda_{\rm L}^{2}} \bigg( \frac{|\Psi|^{2}}{n_{\rm s}} - 1 \bigg)
&=0
,\label{427}
\end{align}
\endgroup  
the equalities in \eqref{421} called the BPS bound are saturated. Note that the existence and uniqueness of $n$ separated BPS vortex solution satisfying the boundary conditions,
$\displaystyle{\lim_{\boldsymbol{x} \to \boldsymbol{x}_{a}} |\Psi| (\boldsymbol{x}) = 0}$ and 
$\displaystyle{\lim_{|\boldsymbol{x}|\to\infty} |\Psi| (\boldsymbol{x}) = \sqrt{n_{\rm s}}}$,
are proved with mathematical rigor \cite{Taubes:1979tm}, in which each arbitrary position $\boldsymbol{x}_{a} = (x_{a}, y_{a})~(a=1,2,\cdots,n)$ stands for $a$-th vortex site.
Topological vortex of vorticity $n$ has the quantized magnetic flux 
$\Phi_{B}=2\pi\Phi_{{\rm L}}n$ for arbitrary $\lambda$ and the inter-distances among vortices while each unit BPS vortex costs the energy $\pi \hbar^{2} n_{\rm s} /m$ and carries the charge $\bar{Q}_{\mathrm{U}(1)} = 4\pi\epsilon_{0} m c^{2} /|q|$.  Since the BPS bound \eqref{421}    means the static $n$ BPS vortices are noninteracting with vanishing stress components of the energy momentum tensor $T_{ij}=0$, it means a nonperturbative balance of static interactions at the critical couplings \eqref{312}--\eqref{327}. In this BPS limit of $\lambda=\lambda_{{\rm c}}$ \eqref{312}, the correlation length $\xi$ becomes equal to the London penetration depth 
$\lambda_{{\rm L}}=\sqrt{\epsilon_{0} m/n_{\rm s}}\,c/|q|$ with the unit Ginzburg-Landau parameter $\kappa=\lambda_{{\rm L}}/\xi$ and the borderline of type I $(0\le\lambda<\lambda_{{\rm c}})$ and I$\!$I $(\lambda>\lambda_{{\rm c}})$ superconductors is reproduced.\footnote{This unity $\kappa=1$ is $\sqrt{2}$ in condensed matter community. The two equal length scales are originated from equal mass of the Higgs and gauge boson in the Bogomolny limit of Abelian Higgs model. The factor $\sqrt{2}$ comes from $1/2$ in front of spatial covariant derivative term of the Lagrangian density \eqref{201}, obtained by taking nonrelativistic limit of the relativistic complex scalar field $\phi= e^{-\frac{i}{\hbar} m c^{2} t} \Psi$.}

\begin{figure}[h]
\centering
\subfigure[]{\includegraphics[width=0.48\textwidth]{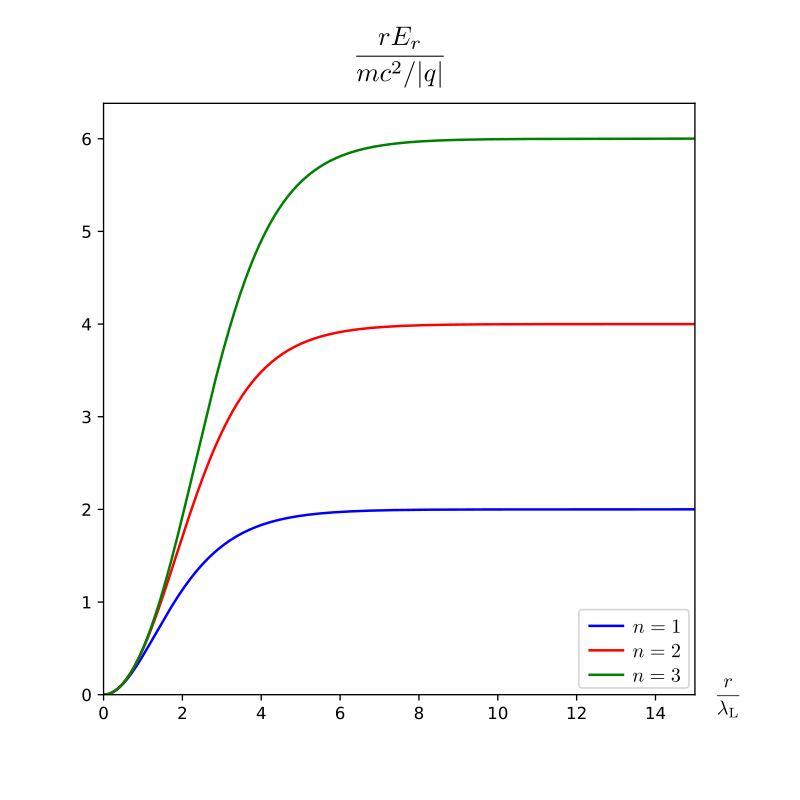}}
\hfill
\subfigure[]{\includegraphics[width=0.48\textwidth]{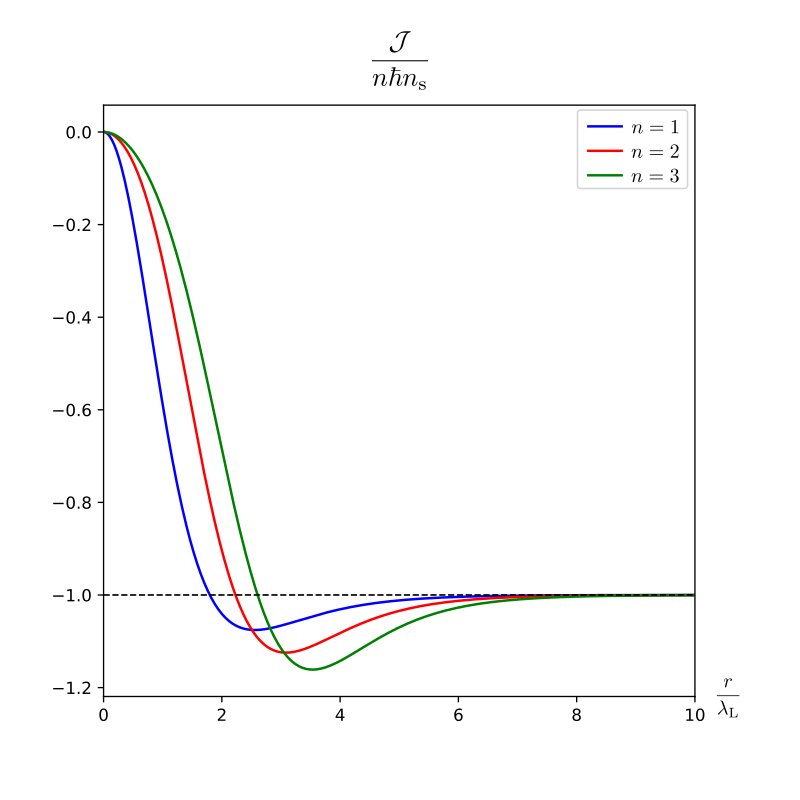}}
\caption{(a) Profiles of $r E_{r}$ in the unit of $mc^{2}/|q|$  for $n=1,2,3$ with $\lambda/\lambda_{\rm c}=1$, as functions of a dimensionless length parameter $r/\lambda_{\rm L}$. Asymptotic value of each curve gives the value of charge proportional to vorticity $n$ in the BPS limit. (b) Profiles of the angular momentum density $\mathcal{J}$ in the unit of $n \hbar n_{\rm s}$.}
\label{fig:1}
\end{figure}

The obtained vortices are distinguished from the electrically neutral 
Abrikosov-Nielsen-Olesen vortices \cite{Abrikosov:1956sx,Nielsen:1973cs} by the nonzero U(1) charge in the superconductivity of $s$-wave. Any static object of nontrivial amplitude of the Schr{\"o}dinger type complex scalar matter, 
$|\Psi|(\boldsymbol{x}) \neq \sqrt{n_{\rm s}}~\,\text{for some planar positions }\boldsymbol{x}$,
gives a nonzero localized charge distribution and the Gauss' law dictates nonzero electric field, $\boldsymbol{E} \neq 0$,
\begin{align}
\boldsymbol{\nabla} \cdot \boldsymbol{E} = \frac{\rho}{\epsilon_0}
=\frac{q}{\epsilon_{0}}(|\Psi|^{2} - n_{\rm s}).\label{209}
\end{align}
If the vortex carries finite electric charge per unit length along $z$ axis $\bar{Q}_{\text{U}(1)}$, it leads to nonzero radial component of the electric field at large distance $r$ of $xy$ plane,
$E_{r} = -d\Phi/dr 
\approx(\bar{Q}_{\text{U}(1)}/2 \pi \epsilon_{0})
(1/r)$ as shown in figure \ref{fig:1}--(a).\footnote{The charge density of the Abelian Higgs model with quadratic time-component of covariant derivative term is proportional to the scalar potential, $\rho=\Phi\big[|\Psi|(\boldsymbol{x})^{2} - v^{2}\big]/c$. Even for static vortex of nonzero matter density, $ |\Psi|(\boldsymbol{x})^{2} - v^{2} \neq 0 $, its U(1) charge density is zero, implemented by the Weyl gauge fixing condition $\Phi=0$.}
Even though Coulombic electric field is obtained, it is cancelled by the gapless neutral scalar field of acoustic phonon for the critical phonon coupling \eqref{327}. It is confirmed by a Bogomolny equation \eqref{422}. The vortex has both electric and magnetic field, and its angular momentum gives a divergent term proportional to the $xy$-planar area of the sample and the vorticity $n$ due to constant background charge density $qn_{{\rm s}}$. However, finite piece of the angular momentum is calculated to be zero for cylindrically symmetric configurations by use of the Gauss' law,
\begin{align}
J + \hbar n_{\rm s} n \int d^{2} \boldsymbol{x}
&=
2\pi \hbar \int_{0}^{\infty} dr\, r 
\bigg[ \frac{\epsilon_{0}}{q} \frac{d\Phi}{dr} \frac{dA}{dr} 
- (A+n) (|\Psi|^{2} - n_{\rm s}) \bigg]
= 0.
\label{431}
\end{align} 
This suggests  the obtained charged vortices are spinless objects, while nontrivial profiles of angular momentum density \(\mathcal{J}\) in figure \ref{fig:1}--(b) say that  $\mathcal{J}$ of each charged vortex consists of two inner and outer regions which are spinning in opposite directions above and below the constant background level $\hbar n_{{\rm s}}n$.

The critical coupling of cubic Yukawa type interaction \eqref{327} is not only for the BPS limit but also for the existence of nonsingular charged vortices of arbitrary $\lambda$. Similar to the equation of scalar amplitude \eqref{463} of $g=0$, the introduction of neutral scalar field of arbitrary $g$ leaves the terms unchanged except the last term in the equation of scalar amplitude. For time-independent configurations, the linear equation of the neutral scalar field $N$ is exactly solved and it is related to the scalar potential, $\Phi=-q\Phi/\epsilon_{0}g+f$. Since the residual gauge degree of freedom $f$ obeying $\nabla^{2} f(\boldsymbol{x})=0$ can always set to be zero, the last term in the equation of scalar amplitude becomes
\begin{align}
- \frac{\epsilon_{0}}{4q\lambda n_{\rm s} \xi^{2}} \Big( g^{2}
- \frac{q^2}{\epsilon_{0}} \Big) \Phi |\Psi| 
.
\label{407} 
\end{align}
According to the same argument of $g=0$ case, no nonsingular vortex solution is supported unless its coefficient $g^{2}-q^{2}/\epsilon_{0}$ vanishes at the critical phonon coupling \eqref{327}. Therefore, it seems that the condition of $g=g_{{\rm c}}$  provides a necessary condition for superconductivity, at least that of conventional superconductivity with vortices.

In this work, we proposed an effective field theory of the Lagrangian density \eqref{201} and find the charged vortices and their BPS limit in addition to the constant neutral superconducting vacuum of zero classical energy. Meanwhile, charged vortices have long history, e.g., \cite{Careri:1965}~$\sim$\cite{Sahu:2022}, which are usually made from the neutral Abrikosov-Nielsen-Olesen vortices by trapping charges. In addition, intriguing theoretical and experimental studies have been made mostly in high $T_{{\rm c}}$ superconductors~\cite{Khomskii:1995, Blatter:1996, Hagen:1991, Kumagai:2001}. Since our result opens a theoretical window for the generic charged vortices in the superconductivity of $s$-wave, it awaits experimental detection and further theoretical studies. 
 The proposed effective field theory with a gapless neutral scalar field of acoustic phonon seems to be the viable effective field theory of nonrelativistic Schr\"{o}dinger type complex scalar field of Cooper pair for field-theoretic description of $s$-wave superconductivity including time dependence, regular vortex solutions, and types I and I$\!$I  \cite{Polchinski:1992ed,Shankar:1993pf,Nagaosa:1999ud,tinkham2004introduction, Wen:2004ym, Bennemann:2008, Coleman:2015,Burgess:2020tbq, Arovas:2019}. 

Although our current analysis is mostly limited to the BPS case, the charged vortices are worth studying for arbitrary $\lambda$, particularly in type I$\!$I superconductors.   In this work, the proposed effective field theory is studied in nonperturbative semiclassical regime, while it should also be tested in perturbative regime, e.g., physical meaning of the obtained critical couplings \eqref{312}--\eqref{327} as force balances and possible inclusion of the gapless neutral scalar field as the residual low energy degrees of acoustic phonon. The Lagrangian of our interest \eqref{201} contains constant background charge density $qn_{{\rm s}}$ and this homogeneous impurity can naturally become inhomogeneous $n_{{\rm s}}\rightarrow n_{{\rm s}}(\boldsymbol{x})$ in the framework of field theory \cite{Hook:2013yda}. Furthermore, a magnetic impurity term can be added and the BPS structure for inhomogeneous Abrikosov-Nielsen-Olesen vortices is sustained in the Abelian Higgs model \cite{Tong:2013iqa}. In the presence of both constant and delta function type impurities, the vacuum and vortices become inhomogeneous and can be interpreted as impurity-vortex composites
\cite{Ashcroft:2018gkp,Kim:2024gfn,Jeon:2025snd} and the study of charged objects in the inhomogeneous version of this effective field theory will be reported elsewhere \cite{NRIAH}.

\section*{Acknowledgement}

The authors would like appreciate Kwang-Yong Choi, Chanyong Hwang, Chanju Kim, O-Kab Kwon, Hyunwoo Lee, Tae-Ho Park, and  D. D. Tolla for discussions on various topics of condensed matter physics and field theory. This work was supported by the National Research Foundation of Korea(NRF) grant with grant number NRF-2022R1F1A1073053 and RS-2019-NR040081 (Y.K.).


\begin{thebibliography}{99}


\bibitem{Ginzburg:1950sr}
V.~L.~Ginzburg and L.~D.~Landau,
Zh. Eksp. Teor. Fiz. \textbf{20}, 1064-1082 (1950)
doi:10.1016/b978-0-08-010586-4.50078-x.

\bibitem{Abrikosov:1956sx}
A.~A.~Abrikosov,
Sov. Phys. JETP \textbf{5}, 1174-1182 (1957).

\bibitem{Polchinski:1992ed}
J.~Polchinski,
[arXiv:hep-th/9210046 [hep-th]].

\bibitem{Shankar:1993pf}
R.~Shankar,
Rev. Mod. Phys. \textbf{66}, 129-192 (1994)
doi:10.1103/RevModPhys.66.129
[arXiv:cond-mat/9307009 [cond-mat]].

\bibitem{Nagaosa:1999ud}
N.~Nagaosa,
\textit{Quantum Field Theory in Condensed Matter Physics},
Springer, 1999,
ISBN 978-3-540-65537-4, 978-3-642-08485-0, 978-3-662-03774-4
doi:10.1007/978-3-662-03774-4.

\bibitem{tinkham2004introduction}
M. Tinkham, \textit{Introduction to Superconductivity}, Dover (2004).

\bibitem{Wen:2004ym}
X.~G.~Wen,
\textit{Quantum Field Theory of Many-Body Systems},
ISBN 978-0199227259
doi.org/10.1093/acprof:oso/9780199227259.001.0001.

\bibitem{Bennemann:2008}
K.~H.~Bennemann and J.~B.~Ketterson,
\textit{Superconductivity Volume 1: Conventional and Unconventional 
Superconductors Volume 2: Novel Superconductors}
ISBN 978-3-540-73252-5,
doi.org/10.1007/978-3-540-73253-2.

\bibitem{Coleman:2015}
P.~Coleman,
\textit{Introduction to Many-Body Physics},
ISBN 978-0521864886
doi.org/10.1017/CBO9781139020916.

\bibitem{Burgess:2020tbq}
C.~P.~Burgess,
\textit{Introduction to Effective Field Theory,}
Cambridge University Press, 2020,
ISBN 978-1-139-04804-0, 978-0-521-19547-8
doi:10.1017/9781139048040.

\bibitem{Arovas:2019}
D.~Arovas and C. Wu,
https://courses.physics.ucsd.edu/2014/Spring/physics239\\
/LECTURES/SUPERCONDUCTIVITY.pdf,
\textit{Lecture Notes on Superconductivity}.

\bibitem{Nielsen:1973cs}
H.~B.~Nielsen and P.~Olesen,
Nucl. Phys. B \textbf{61}, 45-61 (1973)
doi:10.1016/0550-3213(73)90350-7.

\bibitem{Bogomolny:1975de}
E.~B.~Bogomolny,
Sov. J. Nucl. Phys. \textbf{24}, 449 (1976)
PRINT-76-0543 (LANDAU-INST.).

\bibitem{Kim:2024gfn}
Y.~Kim, S.~Jeon, O.~K.~Kwon, H.~Song and C.~Kim,
Phys. Rev. D \textbf{111}, no.4, 045018 (2025)
doi:10.1103/PhysRevD.111.045018
[arXiv:2409.12451 [hep-th]].

\bibitem{Jeon:2025snd}
S.~Jeon, Y.~Kim and H.~Song,
[arXiv:2501.02500 [hep-th]].

\bibitem{Taubes:1979tm}
C.~H.~Taubes,
Commun. Math. Phys. \textbf{72}, 277-292 (1980)
doi:10.1007/BF01197552.

\bibitem{Careri:1965}
G.~Careri, S.~Cunsolo, P.~Mazzoldi and M.~Santini, Phys. Rev. Lett., 15, 392 (1965).

\bibitem{Sahu:2022}
S.~K.~Sahu, S.~Mandal, S.~Ghosh, M.~M.~Deshmukh and V.~Singh
Nano. Lett. 2022, 22, 4, 1665-1671 (2022)
doi:10.1021/acs.nanolett.1c04688.

\bibitem{Khomskii:1995}
D.~Khomskii and A.~Freimuth, Phys. Rev. Lett., 75, 1384–1386 (1995). 

\bibitem{Blatter:1996}
G.~Blatter, M.~Feigel'man, V.~Geshkenbein, A.~Larkin and A.~Otterlo, Phys. Rev. Lett., 77, 566 (1996).

\bibitem{Hagen:1991}
S.~J.~Hagen, C.~J.~Lobb, R.~L.~Greene and M.~Eddy, Phys. Rev. B., 43, 6246-6248 (1991).

\bibitem{Kumagai:2001}
K.~Kumagai, K.~Nozaki and Y.~Matsuda,
Phys. Rev. B \textbf{63}, 114502 (2001)
doi:10.1103/PhysRevB.63.114502.

\bibitem{Hook:2013yda}
A.~Hook, S.~Kachru and G.~Torroba,
JHEP \textbf{11}, 004 (2013)
doi:10.1007/JHEP11(2013)004
[arXiv:1308.4416 [hep-th]].

\bibitem{Tong:2013iqa}
D.~Tong and K.~Wong,
JHEP \textbf{01}, 090 (2014)
doi:10.1007/JHEP01(2014)090
[arXiv:1309.2644 [hep-th]].

\bibitem{Ashcroft:2018gkp}
J.~Ashcroft and S.~Krusch,
Phys. Rev. D \textbf{101}, no.2, 025004 (2020)
doi:10.1103/PhysRevD.101.025004
[arXiv:1808.07441 [hep-th]].

\bibitem{NRIAH}
Y.~Kim, S.~Jeon, and H.~Song, in preparation.

\end{thebibliography}
\end{document}